\documentclass[conference]{IEEEtran}
\usepackage[T1]{fontenc}

\AtBeginDocument{%
 \providecommand\BibTeX{{%
   \normalfont B\kern-0.5em{\scshape i\kern-0.25em b}\kern-0.8em\TeX}}}
    
\usepackage{xspace}

\usepackage{tikz}
\usetikzlibrary{arrows.meta,positioning,fit,shapes.geometric}

\usepackage{booktabs}
\usepackage{siunitx}
\usepackage{array}
\usepackage{float}
\usepackage{adjustbox}
\usepackage{algorithm}
\usepackage{algpseudocode}

\usepackage{amsmath}
\usepackage{mathtools}

\usepackage{graphicx}

\usepackage{subcaption}

\usepackage{xcolor}

\usepackage[numbered]{bookmark}
\usepackage{soul}
\usepackage{url}

\usepackage{pgfplots}
\pgfplotsset{compat=1.14}

\usepackage{graphicx}
\usepackage{adjustbox}
\usepackage{booktabs}
\usepackage{algorithm}
\usepackage{algpseudocode}

\begin{document}

\title{\huge Predicting Acceptance and Review Effort in Human and Agent Pull Requests}

 \author{
   \IEEEauthorblockN{Kartik Ghanshyambhai Pansuriya,
                     Ehsan Ghorbani,
                     Deepak Singh,
                     Eman Abdullah AlOmar}
   \IEEEauthorblockA{Stevens Institute of Technology, 
                     Hoboken, New Jersey, USA\\
                     Email: \{kpansuri,eghorban,dsingh12,ealomar\}@stevens.edu}
 }

\maketitle
\begin{abstract}

Pull requests (PRs) are a central mechanism for reviewing and integrating code changes in modern software repositories. As AI coding agents begin to submit more code changes alongside human developers, maintainers face a new challenge: deciding which PRs are likely to be accepted and which ones may require substantial review effort. This paper studies whether such outcomes can be estimated at the time a PR is opened, before reviewer discussion, CI feedback, or merge decisions are available.

Using the AIDev dataset, we construct a leakage-aware prediction pipeline for human- and agent-authored PRs. The feature set is limited to submission-time information, including PR text characteristics, metadata, repository context, temporal signals, and lightweight diff statistics. We evaluate classical machine-learning models, including Logistic Regression, Random Forests, Gradient Boosting, Extra Trees, and MLPs, across pooled, human-only, agent-only, and balanced contributor views.

Our results show that acceptance prediction is feasible from early signals: tree-based models achieve F$_1$ scores above 0.95, with textual clarity and metadata among the most influential predictors. Review-effort prediction is more difficult. Comment counts and time-to-merge are only modestly explained by submission-time features, suggesting that reviewer availability, project workflow, and team-specific review practices play a major role. These findings indicate that early PR models can support triage and reviewer prioritization, but should be used as advisory tools rather than automated decision-makers.
\end{abstract}



\pagestyle{plain}

\section{Introduction}

Pull requests (PRs) are the primary mechanism through which developers
propose, review, and integrate changes in collaborative software projects \cite{alomar2025deciphering,alomar2021refactoring,alomar2022code,rahman2014insight,watanabe2025use,wyrich2021bots}.
In platforms such as GitHub, PRs serve not only as code-submission units but
also as discussion spaces where maintainers evaluate correctness,
maintainability, risk, and project fit. Because review capacity is limited,
repositories often face long queues, uneven reviewer workloads, and delays
in merging useful contributions.

Prior work has shown that PR outcomes can often be estimated from
submission-time and repository-level signals, including patch size,
description quality, contributor history, project context, and social
factors~\cite{jiang2021predicting,gousios2014exploratory,tsay2014influence,dey2020which}.
However, most existing PR prediction studies were developed in settings
where contributions were assumed to be human-authored. This assumption is
becoming less stable as AI coding assistants and autonomous coding agents
increasingly generate code changes that enter the same review pipeline as
human submissions.

Recent empirical evidence suggests that agent-authored PRs may differ from
human-authored PRs in structure, review dynamics, and acceptance behavior
\cite{horikawa2025agentic}. AI-generated patches may appear more regular in
formatting or style, but still raise concerns about semantic correctness,
maintainability, and hidden behavioral changes. These differences raise an
important question: do early PR signals remain useful when the review queue
contains both human and agent contributions?

This paper investigates two research questions:

\begin{itemize}
  \item \textbf{RQ1:} How accurately can submission-time PR signals predict
        whether a PR will be accepted or rejected?
  \item \textbf{RQ2:} To what extent can PR metadata and code-level features
        estimate review effort, such as review-comment count and
        time-to-merge?
\end{itemize}

To answer these questions, we use the AIDev dataset~\cite{li2025aidev},
which contains PRs authored by both humans and AI agents. We construct a
leakage-aware modeling pipeline that uses only information observable when
a PR is opened, enabling a realistic early-triage setting. By evaluating
pooled, human-only, agent-only, and balanced contributor views, we examine
not only predictive performance but also whether model behavior remains
stable across contributor types.



\section{Motivation}
\label{Section:Problem}


In practice, PR review is not only a technical process but also a resource-allocation problem. Maintainers must decide which submissions deserve immediate attention, which ones are likely to require deeper inspection, and which changes may create avoidable review overhead. These decisions are difficult because many useful signals are incomplete at submission time: reviewers have not yet commented, CI may not have finished, and the final merge decision is unknown.

This problem becomes more complex in repositories that receive both human-authored and AI-generated PRs. Agent-authored changes may look well structured on the surface, but reviewers may still inspect them carefully for hidden semantic errors, missing tests, or maintainability risks. As a result, predictors developed for traditional human-only PR settings may not behave the same way in a mixed human--agent review pipeline.

\begin{figure}[th]
\centering 
\includegraphics[width=1.0\columnwidth]{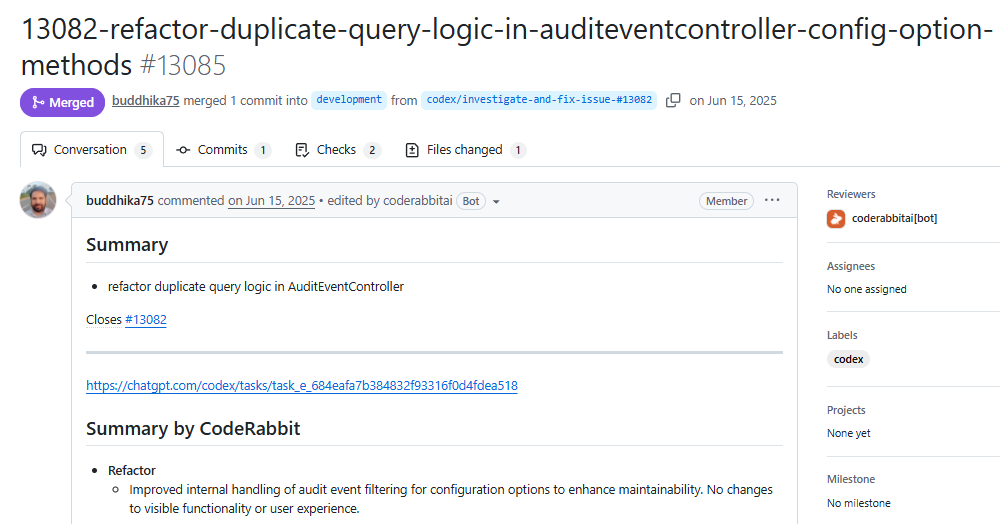}
\caption{Example of an agentic PR \cite{Example}.} 
\label{fig:example}
\vspace{-.2cm}
\end{figure}

The motivation for this study is therefore practical and methodological. Prior work on review latency and PR completion also shows that review
outcomes depend on workflow timing, reviewer availability, automation, and
project activity patterns, not only on the submitted code itself
\cite{yu2015wait,maddila2020nudge,wessel2022github}. Practically, an early prediction model could help maintainers prioritize unclear or potentially costly PRs before review effort accumulates. Methodologically, the mixed human--agent setting allows us to test whether familiar submission-time signals---such as description length, patch size, file churn, and repository context---remain useful when the source of a PR changes. An example of a hybrid review is shown in Figure \ref{fig:example}. 

Our work addresses this gap through four design choices:

\begin{itemize}
  \item \textbf{Submission-time prediction.} We predict PR acceptance (RQ1) and review effort (RQ2) using only information available when the PR is opened, before reviewer discussion, CI outcomes, or merge decisions can influence the model.

  \item \textbf{Mixed human--agent analysis.} We evaluate human-authored and agent-authored PRs both together and separately, allowing us to examine whether predictive patterns remain stable across contributor types.

  \item \textbf{Leakage-aware and balanced evaluation.} We explicitly remove post-review artifacts from the feature matrix and use balanced contributor views to reduce the risk that results are driven only by the dominant PR source.

  \item \textbf{Effort-aware modeling.} In addition to acceptance, we model review effort using comment count and time-to-merge, treating these as operational proxies for reviewer workload.
\end{itemize}

This framing positions early PR prediction as a triage-support problem rather than an automated decision-making task. The goal is not to replace maintainers, but to identify which submissions may deserve earlier attention, additional context, or more careful review in mixed human--AI development environments.

This paper is structured as follows. We start by detailing our methodology in Section \ref{sec:study-design}. Then, we evaluate our approach in Section \ref{sec:experimental-results}. We report in Section \ref{sec:threats} the threats to our work's validity. In Section \ref{sec:related}, we listed the related work, before concluding
and describing our future work in Section \ref{sec:conclusion}.

\section{Study Design}
\label{sec:study-design}

We frame the study as an early-prediction problem over pull requests (PRs). For each PR, the model receives only information that would be visible when the PR is first opened, and it must predict either the final acceptance outcome (RQ1) or a proxy for review effort (RQ2). This setup is intentionally restrictive: it prevents the model from using signals that appear later in the review process and keeps the evaluation close to a real triage scenario.

\begin{figure*}[!t]
    \centering
    \includegraphics[width=\textwidth]{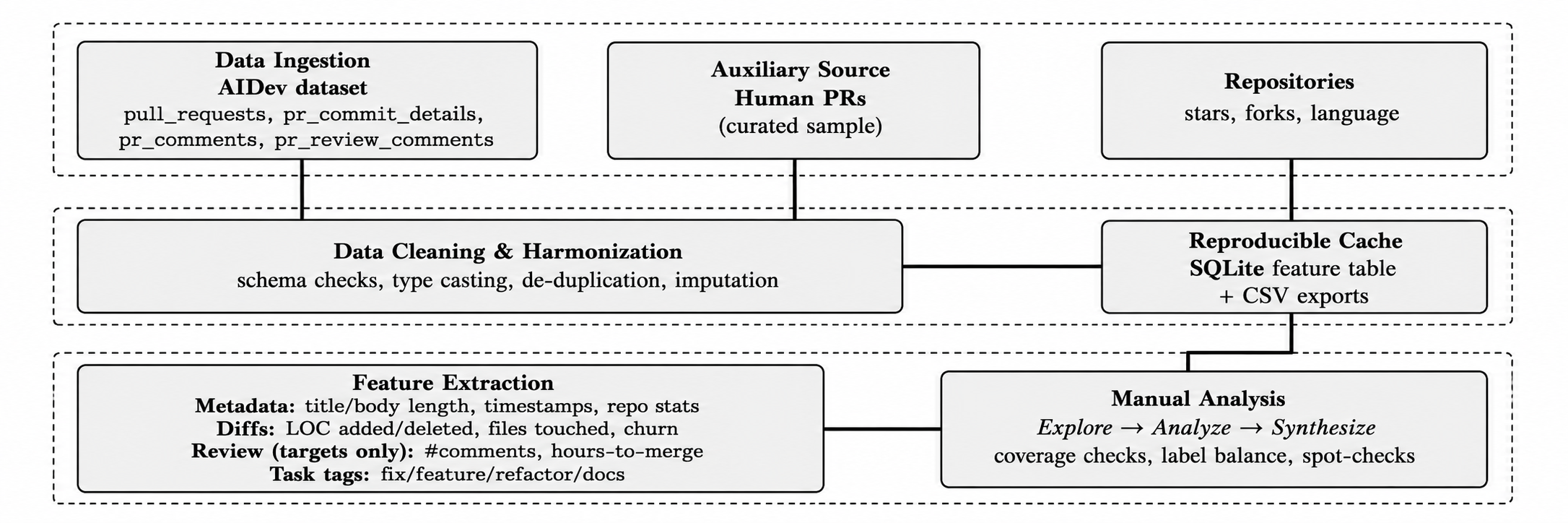}
    \caption{Data preparation and feature engineering workflow: ingestion of AIDev, curated human PRs, and repository metadata; cleaning and harmonization; reproducible caching; feature extraction; and manual validation.}
    \label{fig:data-preparation-workflow}
\end{figure*}

\begin{figure*}[!t]
    \centering
    \includegraphics[width=\textwidth]{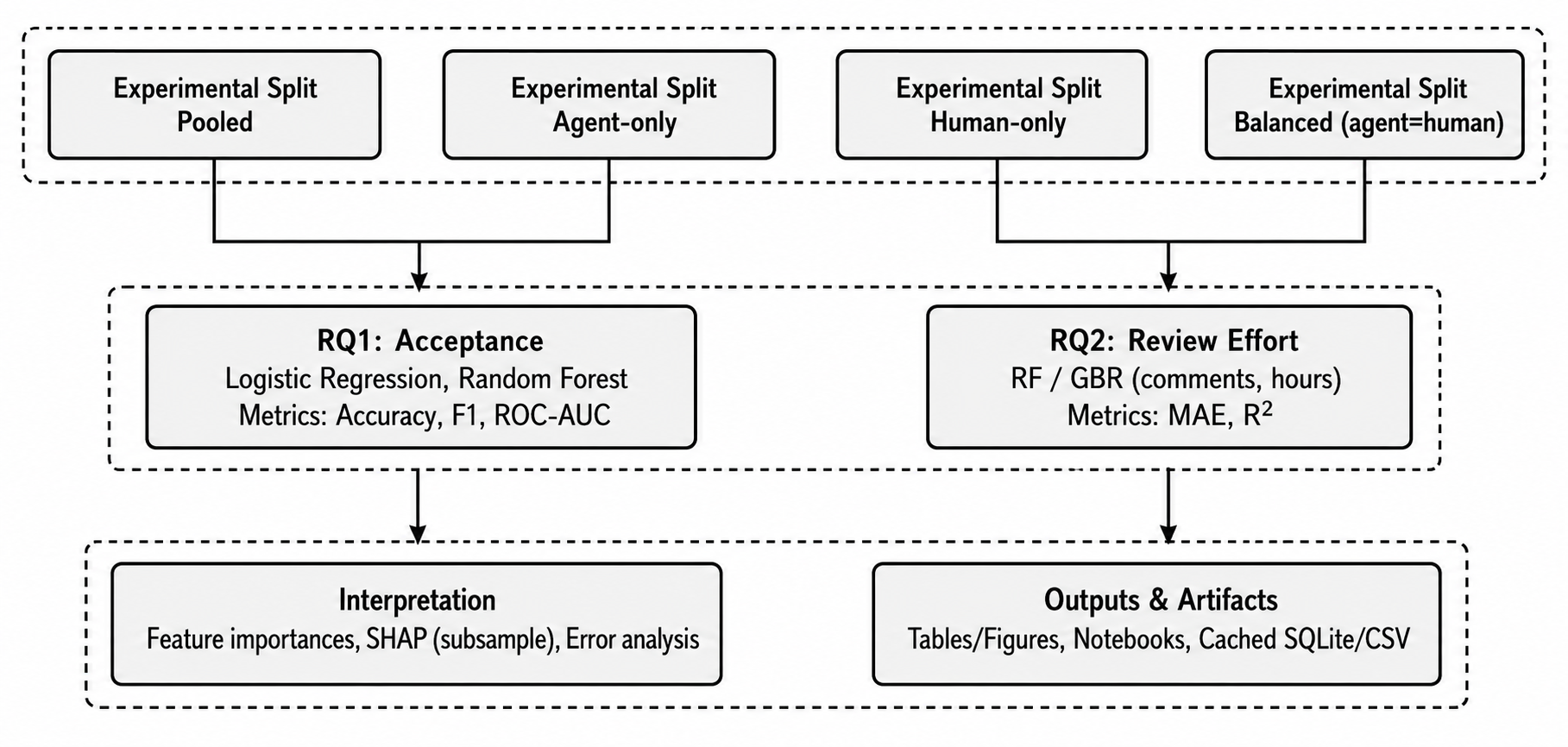}
    \caption{Experimental design and modeling workflow: construction of pooled, agent-only, human-only, and balanced experimental views; RQ1 acceptance classification; RQ2 review-effort regression; and downstream interpretation and artifact generation.}
    \label{fig:experimental-design-modeling}
\end{figure*}

Similar to \cite{watanabe2025use,horikawa2025agentic,horikawa2026ai}, the experiments are based on the \textbf{AIDev} corpus~\cite{li2025aidev}, which contains PR-level metadata, commit-level diffs, review discussions, and repository context for both human- and agent-authored contributions. We use the corpus to build a row-per-PR modeling table and then evaluate whether early signals---including PR text, diff structure, repository metadata, and contributor type---support reliable prediction. Figure~\ref{fig:data-preparation-workflow} summarizes the data preparation and feature-construction workflow, while Figure~\ref{fig:experimental-design-modeling} summarizes the experimental views and modeling pipeline.

\subsection{Objectives and Design Principles}
\label{sec:objectives-principles}

The study is organized around two prediction tasks. \textbf{RQ1} is a binary
classification problem: given a PR at submission time, predict whether it will
eventually be merged or closed without merge. \textbf{RQ2} is a regression
problem: estimate review effort using measurable proxies, specifically
review-comment count and time-to-merge. These tasks are related, but they are
not interchangeable. Acceptance is a final repository decision, while review
effort reflects the amount of interaction and delay that occurs before that
decision is reached.

The design follows five principles. First, all predictors must be observable
when the PR is opened. Second, review-time and post-review artifacts are
excluded from the feature matrix, including review comments, CI outcomes,
review decisions, merge timestamps, and later commits. Third, the joined data
are manually inspected before modeling to check schema coverage, missing
fields, linkage quality, label balance, and extreme values. Fourth, the
experiments are evaluated under pooled, human-only, agent-only, and balanced
views so that performance is not interpreted only from the dominant contributor
group. Fifth, the workflow is made reproducible through deterministic
preprocessing, fixed random seeds, cached feature tables, and exported
experimental artifacts.

\subsection{Data Sources and Schema}
\label{sec:data-sources-schema}

The final modeling table is built at the PR level. Each row corresponds to one
pull request and is assembled from PR metadata, file-level change records,
review-discussion records, and repository information. The main AIDev tables
used in the study are:

\begin{itemize}
  \item \texttt{pull\_requests}: PR identifiers, repository identifiers, title and
  body text, contributor type, state, merge indicator, and creation/update
  timestamps.

  \item \texttt{pr\_commit\_details}: file-level change records, including
  additions, deletions, modified paths, file status, and the number of files
  touched by a PR.

  \item \texttt{pr\_comments} / \texttt{pr\_review\_comments}: discussion and
  review records used to construct RQ2 targets and to assess coverage. These
  tables are not used as input features.

  \item \texttt{repositories}: repository-level context, including stars, forks,
  and primary programming language.
\end{itemize}

Several source tables contain multiple records for the same PR. To avoid
duplicating PRs during joining, we aggregate one-to-many tables before merging.
For example, file-level changes are summarized into PR-level measures such as
total additions, total deletions, churn, file count, and file-type composition.
Review and discussion records are summarized only for target construction and
coverage checks. This approach keeps the unit of analysis consistent: one row
represents one PR. Contributor labels are retained so that the same modeling
pipeline can be evaluated across pooled, human-only, agent-only, and balanced
views.

\subsection{Manual Data Analysis}
\label{sec:manual-data-analysis}

Before model training, we manually inspect the joined dataset to identify
issues that could distort the experiments. This inspection includes checking
table sizes, PR-state labels, missing fields, duplicate identifiers, timestamp
consistency, and whether comment records can be reliably linked back to the
correct PRs. The goal is to verify that the dataset supports the intended
prediction tasks before automated preprocessing begins.

The manual review also separates fields that can be used as predictors from
fields that should only be used as targets or diagnostics. For example,
review comments and merge-related timestamps are useful for defining review
effort, but they occur after PR submission and therefore cannot appear in the
input feature matrix. When unresolved data limitations remain, such as partial
comment coverage or uneven repository representation, we treat them as threats
to validity rather than masking them inside preprocessing.
\subsection{Cleaning and Harmonization}
\label{sec:cleaning-harmonization}

The raw AIDev tables require several harmonization steps before modeling. We
first standardize column names and categorical values across tables. Numeric
fields such as additions, deletions, file counts, stars, and forks are coerced
into consistent numeric formats. Timestamp fields are parsed into a common UTC
format so that creation-time features and time-to-merge targets can be computed
consistently.

Duplicate records are removed using stable identifiers such as repository ID,
PR ID, and creation timestamp. For one-to-many relationships, including changed
files and review comments, records are summarized before joining so that a PR
does not appear multiple times in the final dataset. Missing numeric values are
handled conservatively. Diff counts are treated as zero only when the missing
record indicates no observed change; otherwise, missing values are imputed
within the training fold. Missing categorical values are assigned an explicit
\texttt{unknown} category.

After cleaning, the row-per-PR table is stored in a SQLite cache and mirrored
to CSV and Parquet files. These exports make the workflow easier to rerun and
allow intermediate data products to be inspected when unexpected model behavior
appears.

\subsection{Feature Engineering (Submission-time)}
\label{sec:feature-engineering-study-design}

The feature set is designed to represent information a maintainer could see
when a PR is opened. We do not use deep semantic code analysis in this version
of the study. Instead, we focus on lightweight signals that can be extracted
consistently across repositories and contributor types. The resulting features
fall into four groups.

\textbf{PR metadata and text structure.} From the title and body, we extract
length-based and structure-based indicators, including title length, body
length, whether the body is present, whether URLs appear, and whether the
description includes fenced code blocks. These variables provide simple
proxies for the amount and structure of context supplied by the author.

\textbf{Repository and temporal context.} We include repository-level attributes
such as stars, forks, primary language, and log-transformed popularity
measures. We also extract creation-time indicators such as hours of day and days
of week. These features do not directly measure PR quality, but they help
capture the environment in which the PR enters the review queue.

\textbf{Diff structure.} File-level changes are aggregated into PR-level
measures, including lines added, lines deleted, number of files touched, total
churn, churn ratio, and file-extension diversity. We also compute coarse
file-type mixtures to distinguish code-heavy, documentation-heavy, test-heavy,
and mixed changes. These features approximate the size and spread of a change
without relying on project-specific semantic analysis.

\textbf{Task-intent tags.} Since understanding the intent is crucial \cite{alomar2021we,alomar2019impact}, we assign a lightweight task label from the PR title
and body using keyword rules. The categories include \texttt{fix},
\texttt{feature}, \texttt{refactor}, \texttt{docs}, and \texttt{unknown}. These
tags are intentionally simple. They provide a coarse indication of PR intent
without using reviewer-assigned labels or other post-submission annotations.

For RQ2, review-related quantities, such as time-to-merge and comment count, are
computed only as target variables. They are not included as predictors. This
separation is essential because the purpose of the model is to estimate future
review outcomes from information available at the time of submission.

\subsection{Example of Final Modeling Table}
\label{sec:example-modeling-table}
Table~\ref{tab:sample-prs} shows a small excerpt of the final row-per-PR
modeling table. The sample includes both human- and agent-authored PRs and
illustrates the feature types used in the experiments: text-length indicators,
diff-size measures, acceptance labels, time-to-merge when available, and a
lightweight task tag. The table is included to clarify the data format; it is
not intended to represent the full distribution of the dataset.

\begin{table}[t]
  \centering
  \scriptsize
  \setlength{\tabcolsep}{4pt}
  \renewcommand{\arraystretch}{0.8}
  \caption{Example subset of PRs used in our modeling dataset.}
  \label{tab:sample-prs}
  \begin{adjustbox}{width=1.0\columnwidth,center}
  \begin{tabular}{l l r r r r r r r r l}
    \toprule
    PR ID & Author & Title Len & Body Len & LOC+ & LOC- & \#Files & Churn & Accepted & TTM (hrs) & Task \\
    \midrule
    86   & agent & 30 & 298   & 12  & 4   & 3 & 16  & 1.000 & 0.83 & fix \\
    317  & agent & 35 & 316   & 25  & 10  & 2 & 35  & 1.000 & 1.12 & feature \\
    873  & agent & 49 & 207   & 7   & 2   & 1 & 9   & 1.000 & 0.63 & refactor \\
    315  & human & 56 & 434   & 18  & 6   & 4 & 24  & 1.000 & 3.41 & unknown \\
    6196 & human & 44 & 35602 & 115 & 32  & 7 & 147 & 0.000 & ---  & feature \\
    8440 & human & 23 & 0     & 9   & 1   & 1 & 10  & 1.000 & 0.92 & docs \\
    \bottomrule
  \end{tabular}
  \end{adjustbox}
\end{table}
\subsection{Targets and Leakage Controls}
\label{sec:targets-leakage}

\paragraph{RQ1: Acceptance prediction}
For RQ1, we define PR acceptance as a binary classification task. A PR is
treated as accepted when it is merged into the target branch and rejected when
it is closed without merge. The feature matrix is restricted to information
that would be available when the PR is opened. Variables that reveal later
review activity---including reviewer comments, CI results, later commits,
review decisions, merge timestamps, and final-state artifacts---are excluded
from the predictors.

This restriction is central to the study design. The objective is not to
explain a decision after the review process has already unfolded, but to test
whether an early triage model can provide useful guidance before maintainers
spend review effort.

\paragraph{RQ2: Effort estimation}
For RQ2, we estimate review effort using two observable proxies:
review-comment count and time-to-merge. Time-to-merge is computed only for
merged PRs, since closed PRs do not have a merge timestamp. Comment count is
computed only when discussion or review records can be linked reliably to the
corresponding PR. Both quantities are used strictly as targets; they are not
included as input features.

This task is intentionally harder than RQ1. Acceptance has a clear final
label, while review effort is influenced by factors that are only partly
visible at submission time. A large PR may require little discussion if the
change is routine, while a small PR may take longer if it raises design,
testing, or correctness concerns.
\subsection{Experimental Views and Splits}
\label{sec:experimental-views-splits}

We evaluate the modeling pipeline under four dataset views. The
\emph{pooled} view includes all PRs and reflects the natural mixture of
human- and agent-authored submissions. The \emph{agent-only} view contains
only PRs attributed to coding agents, while the \emph{human-only} view
contains only human-authored PRs. The \emph{balanced} view down-samples the
larger contributor group so that human and agent PRs are represented equally.

These views allow us to examine whether the learned patterns remain stable
across contributor types. Strong performance in the pooled setting alone could
be misleading if it is driven mainly by the dominant contributor group or by
repository-specific distributional effects. The separated and balanced views
therefore provide a check on whether the results reflect general PR-level
signals rather than contributor imbalance.

Within each view, we use 5-fold cross-validation. For RQ1, folds are
stratified by the merged/closed label to preserve the approximate class
distribution. For RQ2, folds are formed over the subset of PRs with the
relevant target available: merged PRs for time-to-merge and reliably linked
PRs for comment-count prediction. Preprocessing, imputation, and
hyperparameter selection are performed only on the training portion of each
fold, and the held-out fold is used only for evaluation.

\subsection{Modeling Pipelines}
\label{sec:modeling-pipelines-study-design}

Both research questions use the same submission-time feature table, but they
differ in learning objective. RQ1 is evaluated as a classification task, while
RQ2 is evaluated as a regression task. We use classical tabular machine
learning models because the feature space is primarily composed of structured
metadata, lightweight text indicators, repository attributes, temporal
signals, and diff statistics. These models are efficient to train, easier to
inspect, and more reproducible than heavier neural approaches, which is
important for a triage-support setting.

For RQ1, we evaluate five classifiers:

\begin{itemize}
  \item \textbf{Logistic Regression (LR)} with $L_2$ regularization and
  standardized numeric features. We use \texttt{class\_weight=balanced} to
  reduce sensitivity to the merged/closed label imbalance.

  \item \textbf{Random Forest (RF)} with 400 trees and depth constraints,
  allowing the model to capture nonlinear interactions among text, repository,
  temporal, and diff-level signals.

  \item \textbf{Gradient Boosting Classifier (GBC)}, which builds an additive
  ensemble of trees and can capture gradual decision boundaries across feature
  combinations.

  \item \textbf{Extra Trees Classifier (ET)}, which introduces randomized split
  selection and provides a complementary ensemble model with a different
  bias--variance profile.

  \item \textbf{Multi-Layer Perceptron (MLP)} with one hidden layer of 128
  units, ReLU activation, and early stopping.
\end{itemize}

We include two baselines to contextualize the learned models. The
\textbf{majority baseline} always predicts the dominant class in the training
fold. The \textbf{heuristic baseline} predicts acceptance using a simple rule
based on PR body length and the number of changed files, with thresholds
selected inside the training fold. These baselines help distinguish meaningful
predictive signal from performance caused by class imbalance or simple
threshold effects.

For RQ2, we train separate regressors for review-comment count and
time-to-merge:

\begin{itemize}
  \item \textbf{Random Forest Regressor (RFR)}, used to model nonlinear
  relationships between submission-time features and effort outcomes.

  \item \textbf{Gradient Boosting Regressor (GBR)}, used as a complementary
  ensemble model for skewed and heavy-tailed effort targets.
\end{itemize}

All models are trained through consistent preprocessing pipelines. Scaling is
applied only when needed, such as for LR and MLP. Tree-based models use the
numeric features directly. Keeping the preprocessing consistent across folds
helps ensure that performance differences reflect model behavior rather than
differences in data handling.

\subsection{Hyperparameter Strategy}
\label{sec:hyperparameter-strategy}

Hyperparameter tuning is intentionally bounded so that the experiments remain
reproducible and easy to rerun. For tree-based models, we search small grids
over the number of estimators, maximum depth, and
\texttt{min\_samples\_leaf}. For the MLP, we tune the hidden-layer size and
learning rate within a predefined range and use early stopping to reduce
overfitting. Logistic Regression uses fixed $L_2$ regularization and serves
as a stable linear reference model.

All tuning is performed within the training portion of each fold. For RQ1,
the model-selection metric is macro-F$_1$, since both accepted and rejected
PRs matter in a triage setting. For RQ2, we tune using $R^2$ and report MAE
and $R^2$, which together describe prediction error and explained variance.
We record fold assignments, random seeds, selected hyperparameters, and
exported outputs so that the experiments can be reproduced.

\subsection{Evaluation Protocol and Metrics}
\label{sec:evaluation-protocol}

For RQ1, we report Accuracy, Precision, Recall, F$_1$, and ROC--AUC.
Accuracy and F$_1$ summarize performance at the selected classification
threshold, while ROC--AUC measures how well the model ranks accepted and
rejected PRs across thresholds. This distinction is important because the
dataset is imbalanced: a model can obtain high F$_1$ by predicting the
dominant merged class frequently, even if its ranking ability is only
moderate.

For RQ2, we report MAE and $R^2$. MAE gives the average prediction error in
the original target units, such as comments or hours. $R^2$ summarizes how
much variance in the target is explained by the model. For comment-count
prediction, we also track target coverage because not every PR has reliably
linked review-discussion records.

All metrics are computed under 5-fold cross-validation. For RQ1, the folds
are stratified by the merged/closed label. For RQ2, evaluation is limited to
PRs where the relevant target is available. We report average fold
performance and use the same split structure across comparable models.

\subsubsection{Statistical Significance Testing}

To assess whether model differences are likely to reflect real performance
differences rather than fold-specific variation, we compare per-fold scores
using paired tests. For RQ1, we compare F$_1$ and ROC--AUC across models.
For RQ2, we compare MAE and $R^2$. Paired tests are appropriate because the
models are evaluated on the same fold partitions. The significance threshold
is set to $\alpha = 0.05$.

\vspace{-0.8em}
\subsection{Interpretability and Error Analysis}
\label{sec:interpretability-error-analysis}

We analyze model behavior beyond aggregate metrics. For global
interpretability, we compute feature-importance rankings from tree-based
models and compare the dominant feature families across experimental views.
We also compute SHAP values on stratified subsamples to inspect local
explanations for selected predictions.

For error analysis, we examine high-loss cases, including false positives and
false negatives for RQ1 and large residuals for RQ2. These cases help
identify situations where submission-time features are insufficient. For
example, a PR may appear clear and low-risk at submission time but later be
rejected because reviewers identify semantic problems, missing tests, or
maintainability concerns. Conversely, a PR may appear risky because it
touches several files, yet still be accepted quickly if the change is routine
within that repository.

We also compare error patterns across human- and agent-authored PRs to check
whether the same feature families drive predictions for both groups. Finally,
we run feature-family ablations by removing text-structure signals,
repository context, diff metrics, or task tags and measuring the resulting
changes in F$_1$, AUC, MAE, and $R^2$.

\vspace{-0.8em}
\subsection{Algorithmic Outline}
\label{sec:algorithmic-outline}

Algorithm~\ref{alg:build-eval} summarizes the full workflow used to construct
the modeling table, evaluate each experimental view, and export diagnostic
artifacts. The important design choice is that preprocessing, tuning, and
evaluation are repeated inside each fold so that held-out data never
influences model fitting.

\begin{algorithm}[H]
\caption{Reproducible Build \& Evaluation}
\label{alg:build-eval}
\begin{algorithmic}[1]
\State Load AIDev tables and normalize identifiers, timestamps, and data types.
\State Aggregate file-level and discussion-level records into one row per PR.
\State Build submission-time features and store the cleaned table in SQLite/CSV.
\State Compute target variables for acceptance, comment count, and time-to-merge.
\For{view $\in$ \{Pooled, Agent, Human, Balanced\}}
  \State Create 5-fold splits using stratification for RQ1.
  \For{fold $=1..5$}
    \State Fit preprocessing steps only on the training split.
    \State Tune model hyperparameters within the training split.
    \State Evaluate the selected model on the held-out fold.
    \State Store predictions, fold metrics, and diagnostic outputs.
  \EndFor
  \State Aggregate metrics and compute feature-importance summaries.
\EndFor
\State Run ablation and SHAP analyses on stratified samples.
\end{algorithmic}
\end{algorithm}

\subsection{Reproducibility and Artifacts}
\label{sec:reproducibility-artifacts}

To support reproducibility, the data-processing and modeling workflow is kept
deterministic. The cleaned row-per-PR table is cached as a SQLite snapshot and
mirrored to CSV and Parquet exports. We also preserve fold assignments,
random seeds, generated figures, result tables, and notebooks used for the
RQ1 and RQ2 experiments. These artifacts allow the experiments to be rerun,
intermediate data products to be inspected, and each reported metric to be
traced back to its corresponding fold and model configuration.

\subsection{Scope Note}
\label{sec:scope-note}

RQ1 is fully implemented across all experimental views and produces stable classification results. RQ2 is more limited. Time-to-merge prediction is evaluated on merged PRs, while comment-count prediction depends on whether review-discussion records can be linked reliably. For that reason, we treat comment-count prediction as an approximate indicator of review intensity rather than a complete measure of reviewer workload.

This scope distinction is important for interpreting the results. Acceptance is a clear final outcome, but review effort is a noisier socio-technical construct. It depends not only on the submitted patch, but also on reviewer availability, project norms, queue pressure, communication style, and workflow timing. The RQ2 results should therefore be read as an early estimate of effort, not as a full explanation of review workload.

\section{Experimental Results}
\label{sec:experimental-results}

This section presents the empirical results for the two prediction tasks. RQ1
examines whether information available when a PR is opened can distinguish
merged PRs from PRs closed without merge. RQ2 examines whether the same
submission-time feature set can provide an early estimate of review effort,
measured through review-comment count and time-to-merge.

All experiments use the cleaned AIDev modeling table described in
Section~\ref{sec:study-design}. Because the goal is early triage, the
evaluation excludes information that becomes available only after review
begins. The reported results therefore reflect what a model could infer at
PR submission time, rather than what could be explained retrospectively after
review activity and merge decisions are known.

\subsection{Experimental Setup}
\label{sec:setup}

The same submission-time feature table is used across all experiments. For
RQ1, the target is the final PR outcome: merged or closed without merge. For
RQ2, the targets are review-comment count and time-to-merge, computed only
for PRs where the corresponding target can be measured reliably.

For acceptance prediction, we compare Logistic Regression, Random Forest,
Gradient Boosting, Extra Trees, and a lightweight MLP. We also include two
baselines: a majority-class predictor and a simple heuristic based on PR body
length and number of changed files. These baselines are included to determine
whether the learned models capture signal beyond class imbalance or simple
threshold-based behavior.

For review-effort estimation, we use Random Forest and Gradient Boosting
regressors. Comment count and time-to-merge are modeled separately because
they represent different aspects of reviewer workload. Comment count reflects
the amount of discussion around a PR, while time-to-merge also depends on
workflow timing, reviewer availability, CI delays, and project-specific
processes.

All reported results are computed using 5-fold cross-validation. Within each
fold, preprocessing, imputation, model fitting, and hyperparameter selection
are performed only on the training split. The held-out split is used only for
evaluation. This prevents information from the test fold from influencing the
model and keeps the evaluation aligned with the leakage-aware design.

\subsection{RQ1: Predicting Pull Request Acceptance}
\label{sec:rq1}

RQ1 asks whether submission-time signals are sufficient to predict whether a
PR will eventually be merged. This task is useful in a triage setting because
maintainers often need to identify which submissions are likely to be routine,
which ones may require closer inspection, and which ones appear uncertain
before review begins.

We train all classifiers on the same feature table and compare them against
the majority and heuristic baselines. This comparison is important because the
dataset is skewed toward merged PRs. In such a setting, high accuracy or
F$_1$ alone can be misleading if a model simply predicts the dominant class.

\begin{figure}[!t]
  \centering
  \vspace{-0.5em}
  \includegraphics[width=0.90\linewidth]{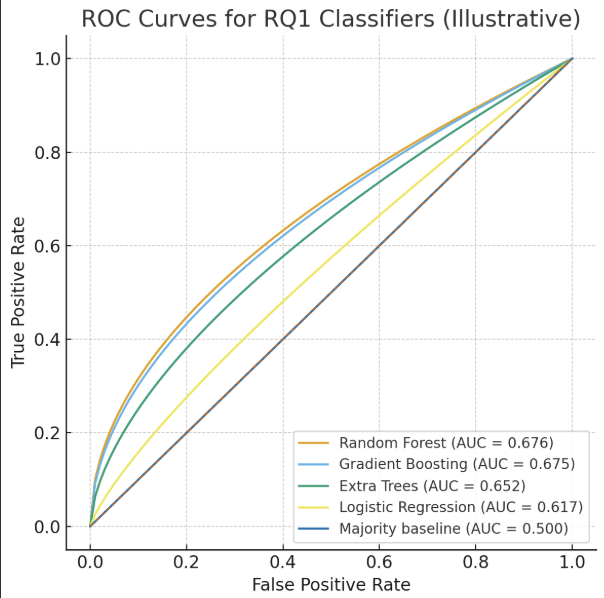}
  \vspace{-0.7em}
  \caption{ROC curves for the main RQ1 classifiers. Tree-based ensembles provide stronger ranking performance than the majority baseline, while Logistic Regression offers a useful calibration reference despite weaker threshold-based performance.}
  \label{fig:rq1_roc}
  \vspace{-0.8em}
\end{figure}

\paragraph{ROC Curves and Calibration}
Figure~\ref{fig:rq1_roc} shows the ROC curves for the main RQ1 classifiers.
The tree-based models, particularly RF and GBC, show stronger ranking
performance than the majority baseline across most threshold settings. This
indicates that acceptance is not captured by a single feature, but by
interactions among PR description quality, repository context, temporal
signals, and diff structure.

The majority baseline remains close to the diagonal, confirming that it has
no meaningful ranking ability even though it performs well under thresholded
metrics. Logistic Regression performs worse than the ensemble models in
classification metrics, but its probability estimates are less extreme. This
makes it a useful calibration reference, even when it is not the strongest
classifier by F$_1$.

\paragraph{Decision Patterns}
The ROC curves suggest that ensemble models are most effective when
identifying lower-risk PRs. These PRs often have enough description context,
moderate change size, and repository signals that make the change easier to
interpret. More difficult cases arise when surface-level signals conflict
with repository-specific context. For example, a PR that touches several
files may appear risky from diff statistics alone, even if the change is
routine within that project.

This pattern shows that acceptance prediction is not merely a patch-size
problem. The strongest models appear to benefit from combining multiple
submission-time cues rather than relying on any single indicator.

\paragraph{Contributor-Type Observations}
Although Figure~\ref{fig:rq1_roc} reports pooled results, we also examine
human-only and agent-only views. Agent-authored PRs show slightly stronger
separability, likely because their formatting and description patterns are
more regular. Human-authored PRs show greater variation in language,
structure, and change scope, which makes their outcomes harder to separate
with a single decision boundary.

Both groups remain predictable from early signals. Agent-authored PRs create
different review dynamics, but their presence does not make submission-time
prediction infeasible.

\subsubsection*{Performance Overview}

Table~\ref{tab:rq1-results} reports the 5-fold cross-validation results for
RQ1. RF, GBC, ET, and MLP form a close performance group, with F$_1$ scores
between $0.957$ and $0.958$. These models outperform the heuristic baseline,
showing that the feature set contains useful acceptance signals beyond simple
rules based on description length or number of changed files.

The majority baseline also achieves a high F$_1$ score because most PRs in
the dataset are merged. However, its AUC of $0.500$ shows that it has no
ability to rank accepted and rejected PRs. This contrast is important: the
learned models are useful for threshold-based triage, but their moderate AUC
values indicate that they should not be treated as perfectly calibrated
probability models.

\subsubsection*{Balanced Contributor View}

To examine contributor-type sensitivity, we repeat RQ1 under a balanced
human--agent view. Logistic Regression improves substantially after
balancing, while RF remains comparatively stable. This suggests that the
linear model is more affected by the original contributor distribution,
whereas the tree-based model is more robust to group imbalance.

\subsubsection*{Feature Importance}

\begin{figure}[!t]
  \centering
  \includegraphics[width=1.0\linewidth]{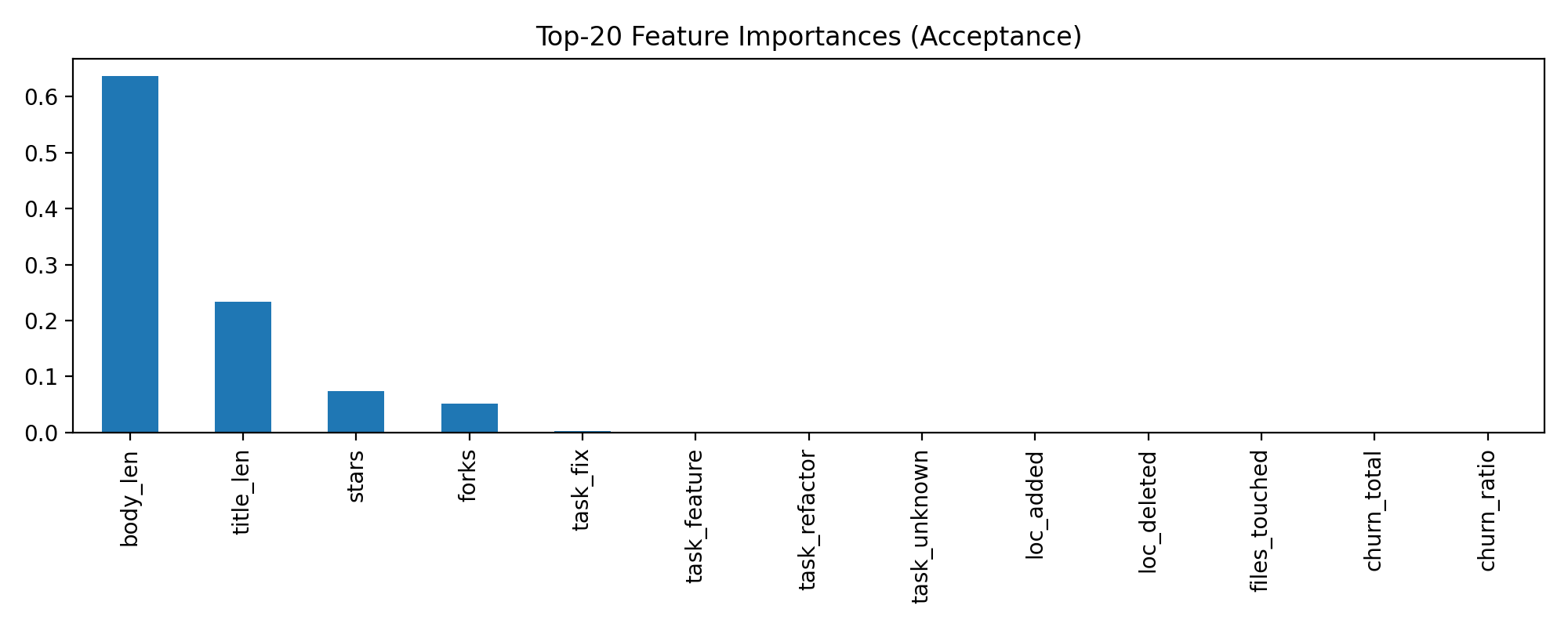}
  \caption{Top-20 feature importances for PR acceptance using Random Forest. Textual and metadata signals rank above most raw diff-size features.}
  \label{fig:feature-importance-rq1}
\end{figure}

Figure~\ref{fig:feature-importance-rq1} reports the top Random Forest
features for acceptance prediction. Text and metadata variables, especially
body length and title length, rank above most raw diff-size measures. This
suggests that how a PR is presented at submission time is strongly associated
with acceptance. These features should be interpreted as predictive signals,
not as causal explanations of reviewer decisions.

\subsection{RQ2: Estimating Review Effort}
\label{sec:rq2}

RQ2 evaluates whether submission-time features can estimate review effort.
We use two targets: review-comment count and time-to-merge. These targets
measure related but distinct aspects of review work. Comment count captures
discussion intensity, while time-to-merge also reflects process delays,
reviewer availability, CI behavior, and repository workflow.

This task is expected to be harder than acceptance prediction. A small PR may
still require substantial discussion if it raises design or correctness
questions, while a large PR may be merged quickly if the change follows an
established pattern. Review effort is therefore better understood as a
socio-technical outcome rather than a property of the patch alone.

\subsubsection*{Regression Results}

Table~\ref{tab:rq2-results} reports the regression results for the two effort
targets. The models capture broad trends, but the explained variance remains
limited. For comment-count prediction, the model reaches an MAE of $1.00$
and an $R^2$ of $0.20$, indicating that submission-time features provide some
early signal about discussion intensity. Time-to-merge is more difficult: the
model obtains an MAE of $24.00$ hours and an $R^2$ of $0.12$.

The lower performance for time-to-merge is consistent with the nature of the
target. Delays can result from reviewer queue length, CI waiting time, team
availability, release timing, or project-specific review practices. These
factors are only weakly captured by the submission-time feature set used in
this study.

\subsubsection*{Feature-Level Patterns}

For RQ2, diff-related variables become more prominent than they were in RQ1.
Lines added, lines deleted, number of files touched, and churn are more
closely tied to review workload than to final acceptance. This pattern is
reasonable: larger or more scattered changes usually require more reading,
more validation, and more discussion.

This differs from the RQ1 results, where text and metadata features are more
influential. In practical terms, acceptance appears more closely linked to how
a PR is framed and contextualized, while review effort is more closely linked
to the amount and spread of change that reviewers must inspect.

\subsubsection*{Contributor-Type Patterns}

Agent-authored PRs show slightly higher discussion and longer median
time-to-merge, consistent with the possibility that reviewers inspect
AI-generated code more cautiously. However, prediction errors are comparable
between human and agent PRs. Thus, agent-authored PRs differ in review
dynamics, but they are not fundamentally less predictable from early signals.

The practical implication is that contributor type should be treated as
context rather than as a direct quality label. An agent-authored PR may
warrant careful inspection, but authorship alone should not determine whether
the PR is treated as risky.

\begin{table}[!t]
\centering
\caption{RQ1 results: 5-fold stratified CV performance across models.}
\label{tab:rq1-results}
\begin{tabular}{lccccc}
\toprule
Model & Acc. & F$_1$ & Prec. & Rec. & AUC \\
\midrule
RF        & \textbf{0.920} & \textbf{0.958} & 0.921 & 0.998 & \textbf{0.676} \\
GBC       & 0.919 & 0.958 & 0.920 & 0.998 & 0.675 \\
ET        & 0.919 & 0.958 & 0.919 & \textbf{1.000} & 0.652 \\
MLP       & 0.919 & 0.957 & 0.919 & 0.999 & 0.652 \\
Majority  & 0.918 & 0.957 & 0.918 & \textbf{1.000} & 0.500 \\
LogReg    & 0.772 & 0.866 & \textbf{0.936} & 0.807 & 0.617 \\
Heuristic & 0.650 & 0.783 & 0.908 & 0.689 & 0.500 \\
\bottomrule
\end{tabular}
\end{table}

\begin{table}[!t]
\centering
\caption{RQ1 balanced contributor view under equal agent--human sampling.}
\label{tab:rq1-balanced}
\begin{tabular}{lccc}
\toprule
Model & Accuracy & F$_1$ & AUC \\
\midrule
LogReg & \textbf{0.873} & \textbf{0.932} & 0.571 \\
RF     & 0.867 & 0.929 & \textbf{0.609} \\
\bottomrule
\end{tabular}
\end{table}
\begin{table}[!t]
\centering
\caption{RQ2 regression results for review-effort prediction.}
\label{tab:rq2-results}
\begin{tabular}{lcc}
\toprule
Target & MAE & $R^2$ \\
\midrule
Review comments & 1.00 & 0.20 \\
Time-to-merge (hours) & 24.00 & 0.12 \\
\bottomrule
\end{tabular}
\end{table}
\subsection{Discussion}
\label{sec:discussion}

The results show that acceptance prediction and review-effort prediction are
not equally difficult. For RQ1, tree-based models reach high F$_1$ scores,
which means that many accepted PRs share recognizable submission-time
patterns. Clearer descriptions, stronger metadata signals, and moderate
change scope appear frequently among PRs that are eventually merged. At the
same time, the moderate AUC values show that these models are not perfect
ranking systems. They are useful for triage, but they should not be used to
automatically decide whether a PR should be accepted.

RQ2 is more difficult because review effort depends on more than the patch
itself. Diff size and churn help estimate broad workload, but they cannot
capture reviewer availability, CI delays, project urgency, or team-specific
review habits. This explains why the regression models only achieve modest
$R^2$ values. The result is still useful: it shows that submission-time
features can provide an early workload signal, but richer workflow and
social context are needed for stronger effort prediction.

The human--agent comparison adds another layer to these findings.
Agent-authored PRs tend to receive more discussion and somewhat longer
review cycles, which suggests that reviewers may inspect AI-generated code
more cautiously. However, prediction errors remain similar across human and
agent PRs. This means that agent authorship changes review dynamics, but it
does not make early prediction unreliable by itself.

Error analysis also shows where early models reach their limit. Some PRs
look strong at submission time because they have clear titles, concise
descriptions, and small diffs, but are later rejected after reviewers find
semantic problems, missing tests, or maintainability concerns. Other PRs
look risky because they touch several files or use vague wording, yet are
accepted quickly because the change is routine in that repository. These
cases show why submission-time models should be used as decision-support
tools rather than replacements for human review.

Overall, the main lesson is that early PR signals are strong enough to
support acceptance triage, but not rich enough to fully explain review
workload. Future systems should combine lightweight submission-time
features with semantic code analysis, CI state, reviewer load, and
author--reviewer history before making stronger claims about review effort.

\section{Related Work}
\label{sec:related}
Dey and Mockus \cite{dey2020which}  designed the probability that a PR will be accepted within a month
after creation using a Random Forest model utilizing 50 predictors. This represents properties of the author, PR, and the project to which
PR is submitted. Their results show an AUC-ROC value of 0.94
with all 14 predictors and 0.77 excluding PR properties that change
after its creation. Gousios et al. \cite{gousios2014exploratory} explored how pull-based software development works, first on the GHTorrent corpus and then on a carefully selected sample of 291 projects. Their finding shows that PR model offers fast turnaround, increased opportunities for community engagement and decreased time to incorporate contributions. Jiang et al. \cite{jiang2021predicting} proposed an accepted PR prediction approach named XGPredict, which builds an XGBoost
classifier based on the training dataset of a project to predict
whether PRs will be accepted. Maddila et al. \cite{maddila2020nudge} designed  an end-to-end service, Nudge, to accelerate overdue pull requests toward completion by reminding the author or the reviewer(s) to engage with their overdue PRs. In a randomized trial involving 147 repositories used at Microsoft, Nudge reduced the PR resolution time by 60\% for 8,500 PRs. This was compared to overdue PRs where Nudge did not send notifications. Rahman and Roy \cite{rahman2014insight}  analyzed PR discussion texts, project-specific information (e.g., domain, maturity), and developer-specific information (e.g., experience) to generate useful insights. These insights are used to compare successful and unsuccessful PRs. Tsay et al. \cite{tsay2014influence} presented a study on open-source software contributions on GitHub that focuses on evaluating PRs. The authors found that project managers used information signaling both good technical contribution practices for a PR and the strength of the social connection between the submitter and the project manager when evaluating PRs. Wessel et al. \cite{wessel2022github} examined how projects use GitHub Actions, what the developers discuss about them, and how project activity indicators change after their adoption. Their results indicate that 1,489 out of 5,000 most popular repositories adopt GitHub Actions and that developers frequently ask for help implementing them. Ren et al. \cite{ren2025hydra} conducted an empirical study to propose a comprehensive taxonomy of code review dimensions. Additionally, the authors identified three major limitations of existing automated code review methods: lack of comprehensiveness, incorrectness, and vagueness. The experimental results show that their tool, Hydra-Reviewer, achieves a BLEU score of 8.20, outperforming the state-of-the-art baseline, DeepSeek-V3, which scores 7.85. Watanabe et al. \cite{watanabe2025use} empirically analyzed 567 GitHub PRs generated with Claude Code, an agentic coding tool, across 157 diverse open-source projects. Their findings indicate that 83.8\% of these agent-assisted PRs are ultimately accepted and merged by project maintainers, with 54.9\% of the merged PRs being integrated without additional changes. AlOmar et al. \cite{alomar2022code}  revealed insights into how reviewers make
a decision about accepting or rejecting a submitted refactoring
request, and what makes such a review challenging. Their main results
report the lack of a proper procedure for developers to follow 
when documenting their refactorings for review, and their survey
with reviewers has also revealed several difficulties related to
understanding the intent of the refactoring.

\section{Threats to Validity}
\label{sec:threats}

The results are based on the AIDev corpus and may not directly generalize
to all open-source or industrial repositories. Review practices vary across
projects, languages, governance models, CI pipelines, and maintainer teams.
The agent-authored PRs in AIDev also reflect the coding agents and workflows
available during the dataset period. As AI coding tools evolve, future
agent-authored PRs may have different structure, error patterns, and review
dynamics. Broader replication across repositories and newer AI-agent
systems is needed before making universal claims.
\section{Conclusion}
\label{sec:conclusion}

This work studied whether submission-time signals can predict pull request
(PR) acceptance and review effort in a mixed human--LLM contributor setting.
Using the AIDev corpus, we built a leakage-aware pipeline that uses only
information available when a PR is opened, including metadata, textual
clarity proxies, repository context, temporal indicators, and lightweight
diff features.

The results show that PR acceptance can be predicted reasonably well from
early signals. Tree-based models achieve strong F$_1$ performance, and the
most useful predictors are largely tied to PR text, metadata, and basic
submission structure. At the same time, the moderate AUC values show that
these models are better suited for triage support than for fully automated
accept/reject decisions.

Review-effort prediction is harder. Diff size, churn, and file count provide
some signal, but comment count and time-to-merge are strongly shaped by
reviewer availability, CI behavior, project workflow, and team-specific
review practices. The human--agent comparison further shows that
agent-authored PRs often receive more scrutiny, while still remaining
predictable from early features when contributor distributions are balanced.

Overall, early PR models can help maintainers prioritize review attention,
but they should augment rather than replace human judgment. Future work
should improve effort prediction by adding semantic code representations,
reviewer--author history, CI state, reviewer workload, and broader
project-level workflow context.

\section{Data Availability}
\label{sec:data-availability}
This replication package containing data/scripts is available online \cite{ReplicationPackage}. 

\section{Acknowledgments} 
 During the preparation of this work, the authors used the ChatGPT Web interface to improve the language and readability. After using this tool, the authors reviewed and edited the content as needed and take full responsibility for the content of the publication. 
\bibliographystyle{abbrv}
\bibliography{sample-base}







\end{document}